\documentclass[a4paper,fleqn,usenatbib]{mnras}

\usepackage[T1]{fontenc}
\usepackage{ae,aecompl}

\usepackage{graphicx}	
\usepackage{amsmath}	
\usepackage{amssymb}	

\newcommand{\Msun}{\mbox{$M_{\odot}$}}
\newcommand{\Rsun}{\mbox{$R_{\odot}$}}

\newcommand{\kms}{\mbox{km s$^{-1}$}}

\title[Magnetic Activity of KIC 12418816]{High Level Magnetic Activity Nature of an Eclipsing Binary KIC 12418816}

\author[H. A. Dal and O. \"Ozdarcan]{H. A. Dal,$^{1}$\thanks{E-mail: ali.dal@ege.edu.tr} and O. \"Ozdarcan,$^{1}$ \\
$^{1}$Ege University, Science Faculty, Department of Astronomy and Space Sciences, 35100 Bornova, \.{I}zmir, Turkey.}

\date{Accepted XXX. Received YYY; in original form ZZZ}

\pubyear{2017}

\begin{document}
\label{firstpage}
\pagerange{\pageref{firstpage}--\pageref{lastpage}}
\maketitle

\begin{abstract}
We present comprehensive spectroscopic and photometric analysis of the detached eclipsing binary 
KIC\,12418816, which is composed of two very similar and young main sequence stars of spectral type 
K0 on a circular orbit. Combining spectroscopic and photometric modelling, we find masses and radii 
of the components as 0.88$\pm$0.06 \Msun\@ and 0.84$\pm$0.05 \Msun\@, and 0.85$\pm$0.02 \Rsun\@, and 
0.84$\pm$0.02 \Rsun\@ for the primary and the secondary, respectively. Both components exhibit narrow emission features superposed on the cores of the \ion{Ca}{ii} H\&K lines, while $H_{\alpha}$ and $H_{\beta}$ photospheric absoprtion is more completely infilled by broader emission. Very high precision $Kepler$ photometry 
reveals remarkable sinusoidal light variation at out-of-eclipse phases, indicating strong spot activity, 
presumably on the surface of the secondary component. Spots on the secondary component appear to migrate towards 
decreasing orbital phases with a migration period of 0.72$\pm$0.05 year.  Besides the sinusoidal variation, 
we detect 81 flares, and find that both components possess flare activity. Our analysis shows that 25 flares among 81 come exhibit very high energies with lower frequency, while the rest of them are very frequent with lower energies.
\end{abstract}




\section{Introduction}\label{S1}

Sinusoidal variations at out-of-eclipses in the light curves of YY\,Gem were observed by 
\citet{Kron_1952ApJ} for the first time in the literature. This unexpected phenomenon, called as 
BY\,Dra syndrome, was explained by \citet{Kunkel_1975IAUS} as a fact that was caused by heterogeneous
temperature distribution, known today as the stellar spot activity on the surface of the star. This
discovery demonstrated that UV\,Ceti type stars also exhibit spot activity. UV\,Ceti stars are 
cool dwarf stars on the main sequence possessing flare activity, which is observed as sudden and 
rapid increase of the stellar flux. The flare activity was discovered during the solar observations by 
R. C. Carrington and R. Hodgson on September 1, 1859 \citep{Carrington_1859MNRAS, Hodgson_1859MNRAS}. 
In the literature lots of studies, such as \citet{2013ApJS20715K}, have been carried out since the first 
flare observation, the general view of the flare activity has been revealed. Nevertheless, 
it has not been yet understood some details of the flare and its process. In general, 
stellar flare activity observed on the dMe stars is modelled on the basis of the solar flare event, 
known as the standard solar flare model \citep{Gershberg_2005}. Both stellar spot and flare activities are 
important in respect to the stellar evolution, because remarkable mass loss is occurred as a results of these 
activities \citep{Benz_Gudel_2010ARA&A}. The red dwarf population rate is about 65\% in our Galaxy, while 
seventy-five percent of them possess flare activity \citep{Rodono_1986NASSP}. 
This means that the population rate of UV\,Ceti type stars in our Galaxy is about 48.75\%. 
However, it is well known for several decades that the general population rate of UV\,Ceti type stars 
is incredibly high in the young stellar clusters, such as the open clusters and the associations 
\citep{Mirzoian_1990IAUS, Pigatto_1990IAUS}. In addition, their population rate decreases
as the cluster age increases. Considering the mass loss caused by stellar flare activity, this situation 
can be explained by the Skumanich law \citep{Skumanich_1972ApJ, Pettersen_1991MmSAI, 
Stauffer_1991ASIC, Marcy_Chen_1992ApJ}. Although several studies and projects on the magnetic activity
occurring on the stars have been accomplished so far, authors have been faced with some
unexplained phenomena. One of them is the active longitude migration. For instance,
\citet{Berdyugina_Usoskin_2003A&A} found two stable active longitudes separated by 180\degr\@ from 
each other on the surface of sun. According to them, these longitudes are exhibiting semi-rigid 
behaviour. On the other hand, these longitudes migrate regularly in time and they are not persistent 
active structures \citep{Arroyo_1961Obs, Stanek_1972SoPh, Bogart_1982SoPh}. The difference between the 
regular activity oscillations of these longitudes, which is so called the flip-flop, is very 
important in terms of the north-south asymmetry exhibited by the magnetic topology of the star. 
Furthermore, calculating angular velocities of these longitudes enlightens the latitudinal rotational 
velocities of spots and spot groups.

Besides the stellar spot activity, another unexplained phenomenon is also observed in case of the stellar 
flares. For instance, it is not absolutely known yet why there are differences between the flare 
energy limits for the stars from different spectral types. In the first place, the highest flare energy detected 
on the Sun increased up to $10^{30}$-$10^{31}$ erg, which is generally obtained from the most powerful 
solar flares, known as two-ribbon flares \citep{Gershberg_2005, Benz_2008LRSP}. The highest level of 
flare energy obtained in case of the Sun is the general value detected for the flares observed in 
RS\,CVn binaries \citep{Haisch_Strong_Rodono_1991ARA&A}. However, flare energy range is a bit larger 
in case of red dwarfs. Detected minimum flare energy is about $10^{28}$ erg, while the observed maximum 
energy is about $10^{34}$ erg for dMe stars \citep{Haisch_Strong_Rodono_1991ARA&A, Gershberg_2005}. 
Apart from all these stars, the highest energy emitting in a flare event is found from the flares of 
the stars in young clusters such as the Pleiades cluster and Orion association. The flare energies 
obtained from these stars can reach $10^{36}$ erg \citep{Gershberg_Shakhovskaia_1983Ap&SS}. 
Considering the standard solar flare model, if there is a difference between the flare energy level, 
similar difference is expected for the other effects of the flare event. In fact, observations 
demonstrated this expectancy, indicating differences between mass loss rates among the stars exhibiting 
the flare activity \citep{Gershberg_2005}. According to the recent studies, the mass loss rate of 
UV\,Ceti type stars is about $10^{-10}$ \Msun\@ per year due to flare like events, while the 
solar mass loss rate is about $2 \times 10^{-14}$ \Msun\@ per year \citep{Gershberg_2005}. The difference 
between the mass loss ratios is also seen between the flare energy levels of the solar and stellar cases. 

In spite of all these differences, flare events occurring on stars from different types are 
generally explained by the standard solar flare model, in which the main energy source is assumed as 
magnetic reconnection process \citep{Gershberg_2005, Hudson_Khan_1996ASPC}. However, to understand the 
whole view of the flare events, parameters, which cause these differences and similarities (among 
singularity, binarity, mass, age, etc.), should be identified. At that point, eclipsing binaries with 
a flaring component are very critical, because physical parameters of a star can be easily determined 
with light curve modelling. In these analyses, the main problem is the initial parameters, especially 
mass ratio of components and surface temperature of the primary component. In addition, to reach the 
real view about flare events, number of samples must be increased. 

KIC\,12418816 was listed in the USNO-A2.0 Catalogue by \citet{Monet_USNO2_0_1998AAS} to the first time 
in the literature. In this catalogue, B and R band magnitudes of the system were listed as 13\fm9 and 
12\fm6, respectively. The brightnesses in the infrared filters $J$, $H$ and $K$ were given as 10\fm872, 
10\fm400 and 10\fm271, respectively \citep{Zacharias_et_al_NOMAD_2004AAS}. The system was classified 
as an eclipsing Algol with a period of 1\fd521925 \citep{Watson_2006SASS}, and with no third light 
contribution. \cite{Coughlin_et_al_2011AJ} tried to determine the parameters of the system for the 
first time in the literature, and they found that the system possesses a light curve with amplitude of 
0\fm581 and an orbital inclination of 87\fdg12. They also found representative effective 
temperature of the system as 4583 K, while the individual temperatures were found as 4603 K and 4563 K 
for the primary and the secondary component, respectively. In the same study, the mass and radius are 
given as 0.78 \Msun\@ and 0.81 \Rsun\@ for the primary component, and 0.77 \Msun\@ and 0.80 \Rsun\@ for 
the secondary component. Moreover, \citet{Slawson_et_al_2011} gave log $g$ and $E(B-V)$ value of the system 
as 4.491 and 0\fm029, respectively. Both \citet{Pinsonneault_et_al_2012ApJS} and \citet{Huber_et_al_2014ApJS} 
confirmed previously given log$g$ value and found $[Fe/H]=-1.51$, while they indicated a bit different 
temperature values compared to the previous studies. More recently, \citet{Armstrong_et_al_2014MNRAS} 
computed the temperature as 4909 K for the primary component and 4796 K for the secondary component, 
which are based on spectral energy distribution fitting. \citet{Morton_et_al_2016ApJ} recently found
an effective temperature of 4998 K with $[Fe/H]=-0.08$, and they estimated the age of the system as 
log(age)=9.53 Gyr in the distance of 175 pc.

In this study, we figure out the nature of KIC\,12418816 via medium resolution spectroscopy and very high
precision space photometry from $Kepler$ spacecraft. In the next section, we summarize source of 
observational data and reduction processes. Section~\ref{S3} comprises spectroscopic and photometric 
modelling of the system, including analysis of spot activity and flares. In the final section, we 
summarize and discuss our findings in the scope of physical properties and magnetic activity nature
of the components.

\section{Observations and data reductions}\label{S2}

\subsection{$Kepler$ photometry}\label{S2.1}

We use detrended and normalized short cadence (58.89 seconds) and long cadence (29.4 minutes) 
$Kepler$ photometry of KIC\,12418816 available at $Kepler$ eclipsing binary catalogue 
\citep{Slawson_et_al_2011, Prsa_et_al_2011}. The catalog includes 42325 data points obtained 
in short cadence photometry, and 57521 data points in long cadence photometry. However, twelfth 
and thirteenth quarters are missing in the long cadence data set, thus we extracted data of 
missing quarters from Mikulski Archive for Space Telescopes (MAST) database. We considered 
simple aperture photometry (SAP) fluxes for these quarters and followed the procedure described 
by \cite{Slawson_et_al_2011} to detrend SAP fluxes. The final long cadence data set includes 
all 17 Quarters of Kepler mission, with 65192 data points in total, and provides almost 
continuous $\sim$4 years of photometry. \cite{Slawson_et_al_2011} estimated one percent 
of contamination due to the other sources close to the KIC\,12418816.

\subsection{Optical spectroscopy}\label{S2.2}

We carried out optical spectroscopic observation of the system with 1.5-m Russian--Turkish 
telescope equipped with Turkish Faint Object Spectrograph Camera 
(TFOSC\footnote{http://www.tug.tubitak.gov.tr/rtt150\textunderscore tfosc.php}) 
at Tubitak National Observatory. 
The instrumental set-up provided medium resolution \'echelle spectra with a resolution of
R = $\lambda/\Delta\lambda$ $\sim$2500 around 6500\,\AA, covering wavelengths between 3900\,\AA\@ 
and 9100\,\AA~in 11 \'echelle orders. All spectra were recorded with a back illuminated 
2048 $\times$ 2048 pixels CCD camera with a pixel size of 15 $\times$ 15 $\mu m^{2}$.

We recorded eleven spectra of our target star between 2014 and 2016 observing seasons. 
Signal-to-noise ratio (SNR) of observed spectra were between 50 and 100 depending on atmospheric 
conditions and exposure time. In addition to target star observations, we obtained high SNR 
optical spectra of 54\,Psc (HD\,3651, K0V), HD\,190404 (K1V) and $\tau$\, Cet (HD\,10700, G8.5V) 
and used these spectra as spectroscopic comparison and radial velocity template.

We followed typical \'echelle spectra reduction steps for reducing observations. We first removed
instrumental noise from all observed frames by using average of nightly obtained bias frames. Then
we obtained average flat-field image from bias corrected halogen lamp frames and normalized the average
flat-field image to the unity. We divided all science and Fe-Ar calibration lamp frames by the normalized
flat-field frame and applied scattered light correction and cosmic ray removal to all flat-field corrected
frames, thus we obtained reduced calibration lamp and science frames. We extracted spectra from reduced
science frames and applied wavelength calibration to them. Finally, we normalized all science spectra
to the unity by using cubic spline function. We applied all reduction steps in IRAF\footnote{The Image
Reduction and Analysis Facility is hosted by the National Optical Astronomy Observatories in Tucson,
Arizona at URL iraf.noao.edu.} environment.

\section{Analysis}\label{S3}

\subsection{Light elements and $O-C$ diagram}\label{S3.1}

We start analysis with determination of mid-eclipse times in long cadence data set. We fit third 
or fourth order polynomial to each eclipse to determine mid-eclipse time. Order of polynomial depends
on the shape and asymmetry of the corresponding eclipse. Then we use determined 
eclipse times to construct an $O-C$ diagram via initial light elements given in $Kepler$ eclipsing
binary catalog (Equation~\ref{Eq1}), and obtain improved light elements by applying a linear fit 
to the $O-C$ data. 

\begin{equation}
    T_{0} {\rm (BJD)} = 2,454,954.742983 + 1\fd5218703 \ \times \ E .
	\label{Eq1}
\end{equation}

Since each eclipse includes only a few data points, polynomial fits to the eclipses leads to lower 
precision, hence a scatter with an amplitude of 0.005 day in $O-C$ diagram. Nevertheless, we continue
with the improved light elements for further spectroscopic orbit and light curve modelling. After we
achieve the best light curve model for the eclipsing binary, we divide the long cadence data into
subsets, where each subset covers only a single orbital cycle, and re-calculate each eclipse time in 
a single cycle of each subset by keeping all light curve model parameters fixed, but only adjusting
ephemeris time. By this way, we are able to determine eclipse times and their statistical errors
reasonably and more precisely. For primary eclipses, this process was pretty straightforward, while 
for the secondary eclipses we needed to change the role of components (thus role of eclipses) in the 
light curve model and shifted each light curve by 0.5 in phase. We repeat this process iteratively until 
we achieve self consistent light elements (Equation~\ref{Eq2}). These light elements are adopted for 
further analysis.

\begin{equation}
    T_{0} {\rm (BJD)} = 2,454,954.74366(3) + 1\fd52187051(5) \ \times \ E .
	\label{Eq2}
\end{equation}

\begin{figure}
\centering
{\includegraphics[angle=0,scale=0.52,clip=true]{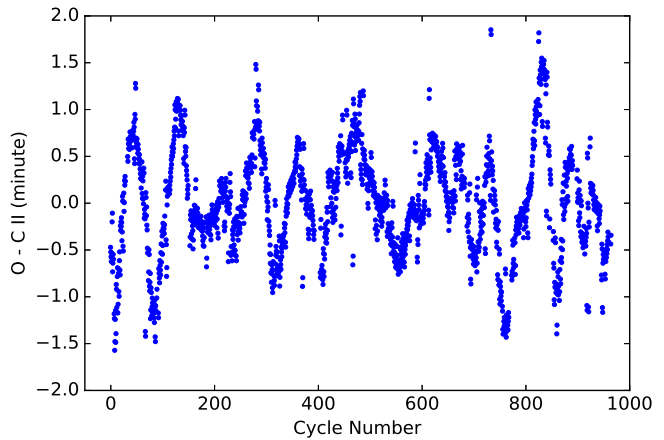}}
\caption{$O-C$ diagram of the eclipse times.}
\label{F1}
\end{figure}

The residuals were derived according to the linear fit obtained from the final eclipse times leads 
to $O-C~II$ diagram shown in Figure~\ref{F1}. There is a wave like variation 
in a form of an irregular sinusoidal wave with an amplitude of 0.001 day. According to the discussions in 
\citet{2015MNRAS.448..429B} and \citet{2013ApJ...774...81T}, the variation is possibly due to the strong 
chromospheric activity of both components (see Section~\ref{S3.5}).

\subsection{Radial velocities and spectroscopic orbit}\label{S3.2}

\begin{figure}
\centering
{\includegraphics[angle=0,scale=0.60,clip=true]{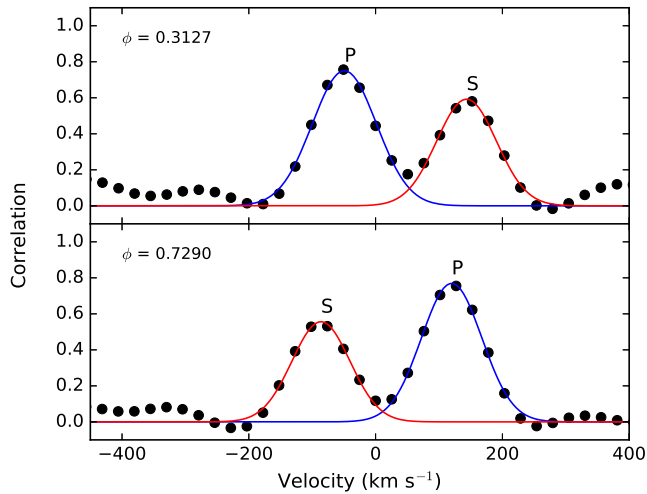}}
\caption{Cross-correlation functions of two spectra obtained at orbital quadratures. The letter $\phi$
denotes corresponding orbital phase. P and S indicate the primary component and the secondary component, 
respectively.}
\label{F2}
\end{figure}

We use 54\,Psc as radial velocity template, and cross-correlate it with each of eleven spectra of
KIC\,12418816 in order to determine radial velocities of the components. We follow cross-correlation
procedure described by \citet{fxcor_Tonry_Davis_1979} via $fxcor$ task under IRAF environment. 
We use absorption lines between 5000\,\AA~and 6500\,\AA, except broad lines (e.g. Na I D lines) and
strongly blended lines, to calculate cross-correlation function. We are able to detect strong and clear
cross-correlation signals from both components. In Figure~\ref{F2}, we plot cross-correlation
functions for two spectra obtained at orbital quadratures.

\begin{table}
\setlength{\tabcolsep}{3pt}
\small
\caption{Brief information on observed spectra, measured radial velocities and their
corresponding standard errors ($\sigma$) in \kms.}\label{T1}
\begin{center}
\begin{tabular}{cccrrrr}
\hline\noalign{\smallskip}
      HJD    & Orbital  & Exposure & \multicolumn{2}{c}{Primary} &  \multicolumn{2}{c}{Secondary} \\
(24 00000+)  &  Phase   & time (s) & V$_{r}$ & $\sigma$ & V$_{r}$ & $\sigma$  \\
\hline\noalign{\smallskip}
56844.4944	&	0.7290	&	3600	&	  96.7	&	5.2	&	-118.1	&	 8.0 \\
56845.3827	&	0.3127	&	3600	&	-111.7	&	5.1	&	  91.0	&	 8.8 \\
56845.4254	&	0.3407	&	3600	&	-101.6	&	6.0	&	  84.3	&	 8.7 \\
56889.2664	&	0.1481	&	2400	&	-103.5	&	7.4	&	  80.3	&	11.0 \\
56890.3491	&	0.8595	&	3600	&	  70.2	&	5.6	&	-100.1	&	 9.7 \\
56890.5051	&	0.9620	&	2400	&	  18.0	&	6.3	&	 -41.9	&	 7.3 \\
57592.3044	&	0.1045	&	1200	&	 -81.4	&	7.3	&	  69.3	&	10.7 \\
57601.3201	&	0.0287	&	3600	&	 -22.0	&	4.0	&     ---   &	---  \\
57601.4964	&	0.1445	&	3600	&	 -97.5	&	6.9	&	  75.1	&	 9.3 \\
57616.4126	&	0.9457	&	3600	&	  20.0	&	7.2	&	 -52.8	&	11.4 \\
57617.4558	&	0.6312	&	3600	&	  69.7	&	5.7	&	 -90.5	&	10.0 \\
\noalign{\smallskip}\hline
\end{tabular}
\end{center}
\end{table}

\begin{figure}
\centering
{\includegraphics[angle=0,scale=0.55,clip=true]{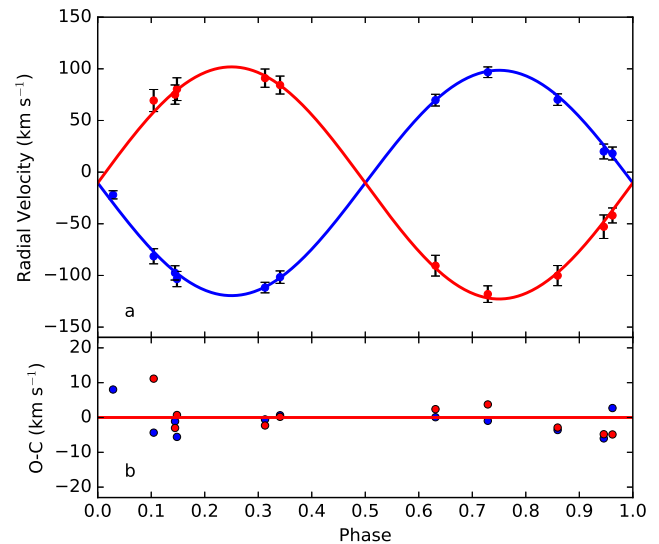}}
\caption{a) Observed radial velocities of the primary and the secondary (blue and red filled
circles, respectively), and their corresponding theoretical representations (blue and red curves).
b) Residuals from theoretical solution.}
\label{F3}
\end{figure}

We tabulate measured radial velocities in Table~\ref{T1}, together with brief information on observed
spectra. Preliminary inspection of long cadence light curve indicate no considerable eccentricity
for the orbit, hence we determine spectroscopic orbit of the system under circular orbit assumption. 
Application of Levenberg-Marquardt algorithm \citep{Levenberg_1944, Marquardt_1963}, and Markov chain 
Monte Carlo simulations via $Python$ package $emcee$ \citep{emcee_Foreman_Mackey_2013PASP} to the measured radial 
velocities and their errors, which is done via a simple script written in $Python$ language, leads 
to the spectroscopic orbit parameters tabulated in Table~\ref{T2}. We plot phase-folded radial velocities 
and theoretical spectroscopic orbit in Figure~\ref{F3}, together with the residuals from the best-fitting 
model.

\begin{table}
\caption{Spectroscopic orbital elements of KIC\,12418816. $M{_1}$ and $M{_2}$
denote the masses of the primary and the secondary component, respectively, while $M$ shows the total
mass of the system.}\label{T2}
\begin{center}
\begin{tabular}{cc}
\hline\noalign{\smallskip}
Parameter & Value \\
\hline\noalign{\smallskip}
$P_{\rm orb}$ (day)     &    1.52187051 (fixed)  \\
$T_{\rm 0}$ (HJD2454+)   &  954.74366 (fixed)  \\
$\gamma$ (\kms)          &    $-10.3\pm1.3$     \\
$K_{1}$ (\kms)           &     108.5$\pm$2.3     \\
$K_{2}$ (\kms)           &    112.9$\pm$3.9     \\
$e$                      &      0 (fixed)       \\
$a\sin i$ (\Rsun)        &     6.66$\pm$0.14    \\
$M\sin^{3} i$ (\Msun)    &    1.710$\pm$0.079  \\
Mass ratio ($q=M{_2}/M{_1}$)   &     0.96$\pm$0.04    \\
rms1 (\kms )             &         4.1         \\
rms2 (\kms )             &         4.6         \\
\noalign{\smallskip}\hline
\end{tabular}
\end{center}
\end{table}

\subsection{Spectral type and features}\label{S3.3}

\begin{figure}
\centering
{\includegraphics[angle=0,scale=0.60,clip=true]{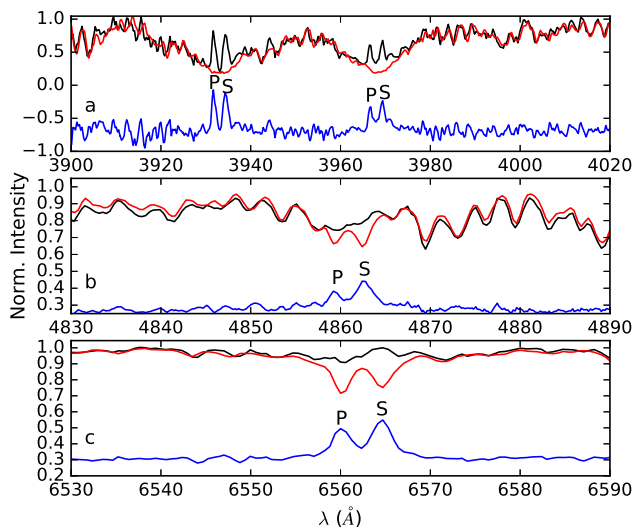}}
\caption{\ion{Ca}{ii} H\&K ($a$), $H_{\beta}$ ($b$) and $H_{\alpha}$ ($c$) regions. Emission in 
$H_{\alpha}$ and $H_{\beta}$ is clearer in residuals from model spectrum. Components are marked 
with P and S, denoting the primary and the secondary component, respectively. Black line shows
the observed spectrum, red line is the model for the composite spectrum (see the text), and blue
line is the difference between the observation and the model, which is shifted upwards by 0.3 for 
the sake of simplicity.}
\label{F4}
\end{figure}

Before proceeding with the spectral classification, we notice exceptional behaviour of Balmer lines 
$H_{\alpha}$ and $H_{\beta}$, which are not in absorption, but filled with emission and embedded 
into the continuum in our medium resolution spectra. There are strong emission features 
in \ion{Ca}{ii} H\&K lines, where we can clearly detect emission from both components. 
In Figure~\ref{F4}, we show these spectral lines.

\begin{figure*}
\centering
{\includegraphics[angle=0,scale=0.80,clip=true]{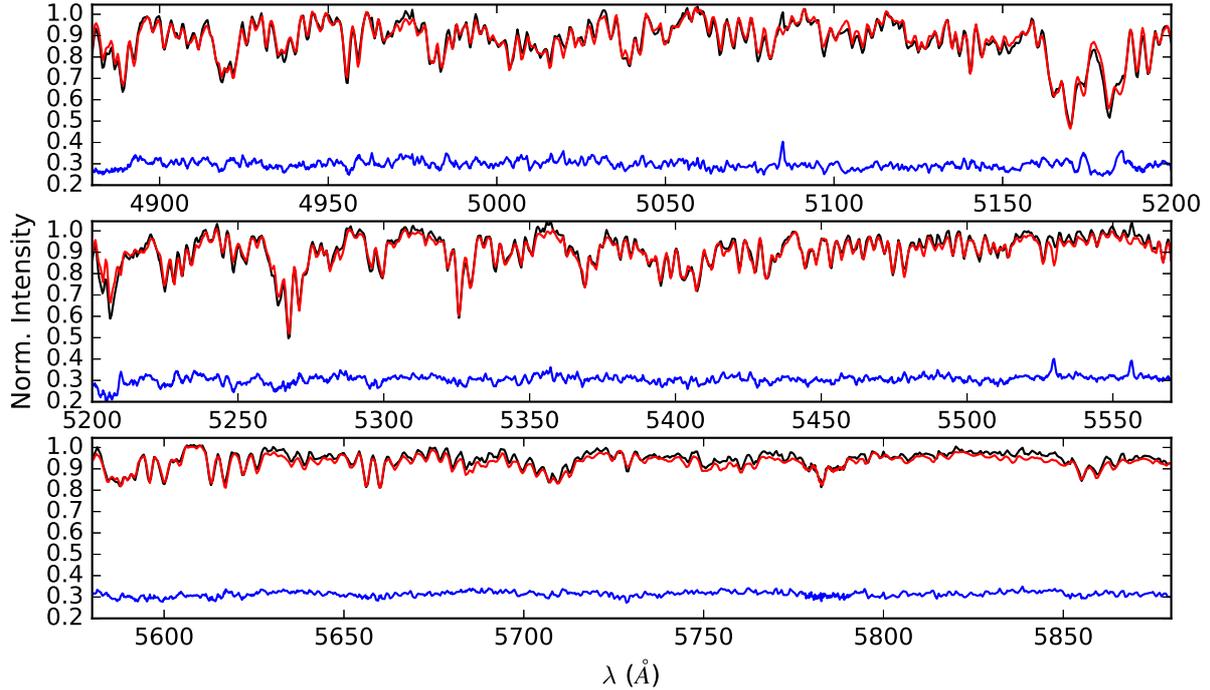}}
\caption{Different portions of selected KIC\,12418816 spectrum. Meaning of the colours are the same as 
in Figure~\ref{F4}. Composite spectrum is for 54\,Psc + 54\,Psc configuration. Residuals between the
selected and composite spectra are shifted upwards by 0.3 for the sake of simplicity.}
\label{F5}
\end{figure*}

We use high SNR spectra of 54\,Psc \citep[K0V,][]{Gray_2003AJ_54Psc}, HD\,190404 
\citep[K1V,][]{Frasca_2009A&A_HD190404} and $\tau$\, Cet \citep[G8.5V,][]{Gray_2006AJ_Tau_Cet} 
as comparison templates to estimate the spectral type of the components. Spectral types of these
stars were reliably determined via high resolution spectroscopic observations in the given 
references. Among eleven spectra of KIC\,12418816, we select the spectrum 
recorded on the night of HJD 24\,56845, where we can clearly separate spectral lines of both 
components. Then, we calculate composite spectrum for each binary combination of template stars, 
(54\,Psc+HD\,190404, HD\,190404+$\tau$\, Cet, HD\,190404+HD\,190404, etc.) and compare each calculated 
composite spectrum with the selected spectrum of KIC\,12418816. For a given
binary combination, we first apply proper radial velocity shift to the spectrum of each template star 
to match their spectral lines to the lines of corresponding component along the wavelength. Then, we
calculate resulting composite spectrum by considering the luminosity ratio of the components that we find
from light curve modelling. We iterate this process until we achieve agreement between spectral types, and
luminosity ratio found from light curve modelling. Among all possible binary configurations of the template stars 
mentioned above,  54\,Psc+54\,Psc configuration with a luminosity ratio ($L_{1}/L_{2}$) of 1.095 
(see Section~\ref{S3.4}) provides fairly good representation of the observed spectrum, and indicates 
K0V spectral type for both components of KIC\,12418816 with solar metallicity. K0V spectral type corresponds
to 5250 K of effective temperature \citep{Gray_2005}. Considering S/N of observed spectra and resolution of 
TFOSC, we estimate the uncertainty in effective temperature as 200 K. In Figure~\ref{F5}, we show different 
portions of selected KIC\,12418816 spectrum, and calculated composite spectrum from the combination of 
54\,Psc + 54\,Psc.

Estimated temperature of the primary component, $T_{1}$, differs more than 250 K from the temperature 
estimated by \citet{Armstrong_et_al_2014MNRAS} (4909 K) and \citet{Morton_et_al_2016ApJ} (4998 K).
\citet{Armstrong_et_al_2014MNRAS} estimated the temperature via spectral energy distribution of the 
star, which is based on photometric measurements in different bands. \citet{Morton_et_al_2016ApJ} 
used 3D linear interpolation in mass-[Fe/H]-age parameter space for a given stellar model grid. 
Since these approaches are not as precise as spectroscopic methods in terms of temperature 
determination, it is not surprising to find considerably different temperature in case of 
KIC\,12418816. Our temperature estimation is based on fitting of observed spectra in a wide 
optical wavelength range, which provides more precise and reliable atmospheric parameters,
compared to the above-mentioned methods.

\subsection{Light curve modelling}\label{S3.4}

\begin{figure}
\centering
{\includegraphics[angle=0,scale=0.50,clip=true]{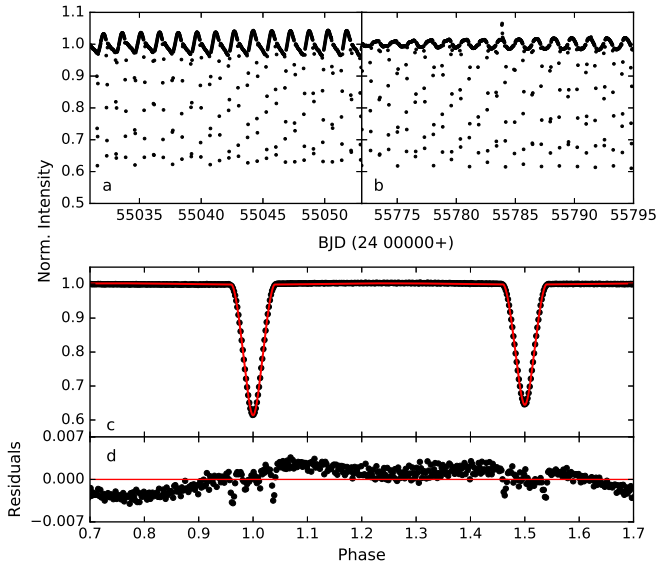}}
\caption{Panel $a$ and $b$ show different parts of long cadence data. Phase binned average light 
curve is in panel $c$ (black filled circles) together with the best-fitting light curve model 
(red curve). Residuals from the model is shown in panel $d$.}
\label{F6}
\end{figure}

We focus on long cadence photometry to obtain radiative and physical properties of KIC\,12418816.
Before modelling, we create phase-binned average light curve of the binary\footnote{Phase binning
is done by $lcbin$ code written by John Southworth and freely available at 
http://www.astro.keele.ac.uk/$\sim$jkt/codes.html$\#$lcbin}
by using light the elements given in Equation~\ref{Eq2} and 65192 long cadence data point. Phase binning
step is 0.002 and the average light curve includes 500 data points. Figure~\ref{F6} panel $a$ and $b$
show different time ranges from long cadence photometry, where considerable differences in the shape
of light maxima, and flares are observed.

We model the phased and binned light curve with 2015 version of the Wilson-Devinney code 
\citep{WD_MAIN_1971ApJ, WD2015_2014ApJ}. The phase-binned average light curve clearly points to
detached configuration, while phases of eclipses (1.0 and 1.5 for the primary and the secondary eclipses, 
respectively) indicate circular orbit for the binary (Figure~\ref{F6} panel $c$, black filled
circles). The most critical two parameters of light curve modelling, i.e. mass ratio ($q=M_{2}/M_{1}$)
and effective temperature of the primary component ($T_{1}$) have already been determined in previous
sections, thus modelling process is pretty straightforward and only requires adjustment in phase shift, 
orbital inclination ($i$), effective temperature of the secondary component ($T_{2}$), dimensionless 
potentials of the components ($\Omega_{1}$ and $\Omega_{2}$) and luminosity of the primary component
($L_{1}$). Albedo ($A_{1}$ and $A_{2}$) and gravity darkening coefficients ($g_{1}$ and $g_{2}$) 
are set to their typical values for convective stars, and square root limb darkening law 
\citep{Sqrt_LD_Law_Klinglesmith_1970AJ} for $Kepler$ passband was adopted, where the limb darkening 
coefficients ($x_{1}$, $x_{2}$, $y_{1}$, $y_{2}$) and bolometric limb darkening coefficients 
($x_{1bol}$, $x_{2bol}$, $y_{1bol}$, $y_{2bol}$) are taken from tables of \citet{van_Hamme_LD_1993AJ}. 
Considering circular orbit, we assume synchronized rotation for the components, thus fixed rotation 
parameter of components ($F_{1}$ and $F_{2}$) to the unity. We tabulate parameters of best-fitting 
light curve model in Table~\ref{T3}, and plot the model in Figure~\ref{F6} panel $c$ with red curve.

\begin{table}
\caption{Parameters of best-fitting light curve model for KIC\,12418816. $\langle r_{1}\rangle$ 
and $\langle r_{2}\rangle$ denote mean fractional radii of the primary and the 
secondary component, respectively. Internal errors of the adjusted parameters 
are given in parentheses for the last digits. Asterisk symbols in the table 
denote fixed value for the corresponding parameter. Note that we adopt the 
uncertainty of $T_{1}$ for $T_{2}$ as well, since the internal error of
$T_{2}$ is unrealistically small ($\sim$2 K).}\label{T3}
\begin{center}
\begin{tabular}{cc}
\hline\noalign{\smallskip}
Parameter & Value \\
\hline\noalign{\smallskip}
$q$ &  0.96* \\
$T_{1}(K)$ &  5250* \\
$g_{1}$, $g_{2}$  & 0.32*, 0.32*\\
$A_{1}$, $A_{2}$  & 0.5*, 0.5*\\
$F_{1}$ = $F_{2}$  & 1.0* \\
phase shift        & $-$0.00003 (2) \\
$i~(^{\circ})$ &  86.53 (1)\\
$T_{2}(K)$ &  5162 (200)\\
$\Omega_{1}$ & 8.805 (31)\\
$\Omega_{2}$ & 8.606 (32) \\
$L_{1}$/($L_{1}$+$L_{2})$ & 0.525 (4) \\
$x{_1bol},x{_2bol}$ & 0.646*, 0.645*\\
$y{_1bol},y{_2bol}$ & 0.177*, 0.172*\\
$x{_1}, x{_2}$  & 0.744*, 0.747* \\
$y{_1}, y{_2}$  & 0.176*, 0.162* \\
$\langle r_{1}\rangle, \langle r_{2}\rangle$ & 0.1277 (5), 0.1266 (5) \\
Model rms           &     6.7 $\times$ 10$^{-3}$   \\
\noalign{\smallskip}\hline
\end{tabular}
\end{center}
\end{table}

\begin{figure}
\centering
{\includegraphics[angle=0,scale=0.50,clip=true]{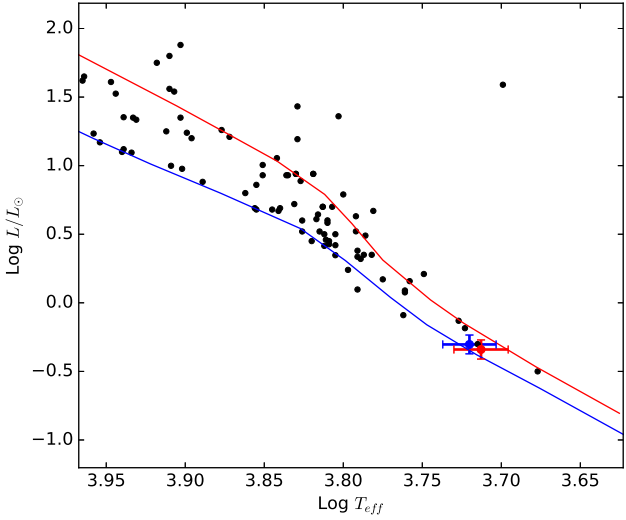}}
\caption{Components of KIC\,12418816 (blue and red filled circles for the primary and 
the secondary component, respectively) in $Log~T_{eff}-Log~L/L_{\odot}$ plane. Blue and
red lines denote ZAMS and TAMS, respectively \citep{Pols_et_al_1998MNRAS}. Black filled 
circles show positions of the components of detached eclipsing binaries, taken from 
\citet{ibanoglu_et_al_2006MNRAS}.}
\label{F7}
\end{figure}

Combined spectroscopic orbit and light curve modelling results yield physical parameters 
of the system listed in Table~\ref{T4}, indicating that the system is composed of two stars
very similar to each other in terms of physical parameters and evolutionary status. 
We plot the components on $Log~T_{eff}-Log~L/L_{\odot}$ plane in Figure~\ref{F7}.

\begin{table}
\caption{Absolute physical properties of KIC\,12418816. Error of each parameter is 
given in paranthesis for the last digits.}\label{T4}
\begin{center}
\begin{tabular}{ccc}
\hline\noalign{\smallskip}
Parameter & Primary & Secondary \\
\hline\noalign{\smallskip}
Spectral Type      &  K0V     &  K0V  \\
\multicolumn{1}{c}{[Fe/H]} & \multicolumn{2}{c}{0.0} \\
Mass (\Msun)       &  0.88(6) & 0.84(5) \\
Radius (\Rsun)     &  0.85(2) & 0.84(2) \\
Log $L/L_{\odot}$   &  $-$0.303(68) & $-$0.340(69) \\
log $g$ (cgs)      &  4.521(15) & 4.511(9) \\
$M_{bol}$ (mag)      &  5.51(17) & 5.60(17) \\
\noalign{\smallskip}\hline
\end{tabular}
\end{center}
\end{table}

\subsection{Stellar cool spot activity}\label{S3.5}

\begin{figure}
\centering
{\includegraphics[angle=0,scale=0.43,clip=true]{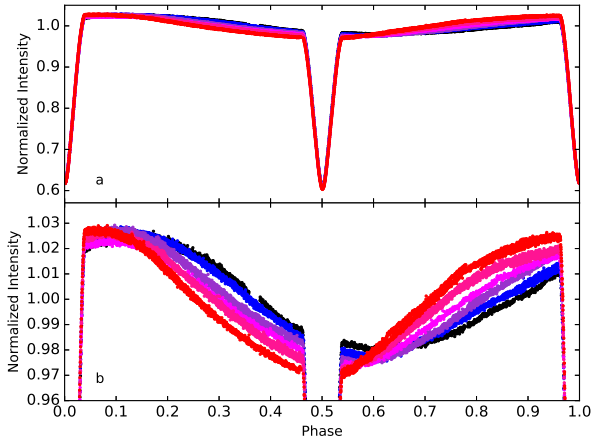}}
\caption{a) The light curves obtained from the available short cadence data are plotted 
versus phase. b) The same light curve is plotted by expanding the intensity axis to show 
the sinusoidal variations clearly.}
\label{F8}
\end{figure}

The available data indicate that the system exhibits also sinusoidal variations at 
out-of-eclipses in light curves. The sinusoidal variation is so distinctive that it is easily 
noticed in whole light curve, in spite of occasional flares and much deeper eclipses. 
Considering the surface temperatures of the components of the system, it is possible that 
the sinusoidal variation must be caused by the rotational modulation due to the cool stellar 
spots. Short cadence data plotted in Figure~\ref{F8} clearly shows that the shape and amplitude 
of the sinusoidal variation are absolutely changed from one cycle to the next one. This situation 
indicates that the active regions on the components are rapidly evolving and also migrating on the 
surfaces of the components.

Before examining out-of-eclipse variations, we first removed the best-fitting eclipsing binary model
from the long cadence data. In practice, we follow the method described in Section~\ref{S3.1}, which
was used to determine precise eclipse times. In the method, we obtain residuals from the theoretical
model, after fitting the ephemeris time. This procedure eliminates any shift in the ephemeris time, 
which could arise from any physical reason, such as third body or spot activity, thus provides
residuals precisely. After obtaining the residuals from whole long cadence data, we then removed
all visually-determined large flares from the residuals. We use the final residuals for further analysis in this section.
Finally, we fit third or fourth degree polynomial to the data points around the deepest light minimum in 
an orbital cycle to determine the corresponding time of that minimum. As in mentioned in Section~\ref{S3.1}
order of polynomial is chosen according to the shape and asymmetry of the light curve for a given cycle. 
After determining the minima times, we calculate orbital phase of each minimum using the light elements 
given in Equation~\ref{Eq2}.

\begin{figure}
\centering
{\includegraphics[angle=0,scale=0.46,clip=true]{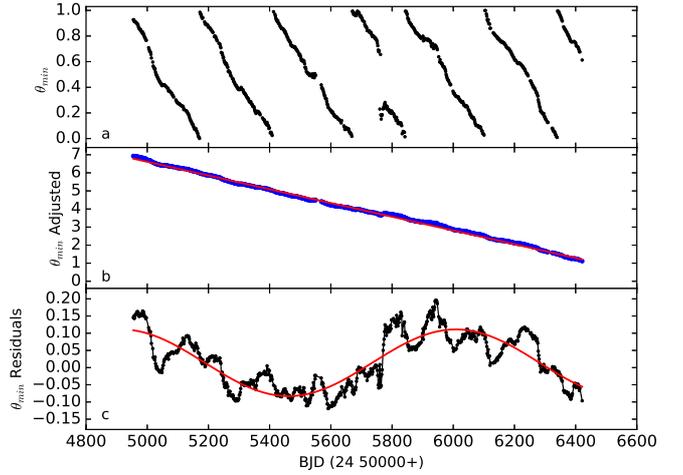}}
\caption{a) The phases of minima times of the sinusoidal variations ($\theta_{min}$) are plotted versus 
the Barycentric Julian Date. b) Similar to $a$ but for adjusted phases. Red line shows linear
fit to the adjusted phases. c) $\theta_{min}$ residuals derived according to the linear fit 
given in Equation~\ref{Eq3}. The red curve represents a sine wave obtained as the best fit 
for the residuals.}
\label{F9}
\end{figure}

In Figure~\ref{F9} panel $a$, we plot the calculated phases of minima (hereafter $\theta_{min}$) versus 
Barycentric Julian Date. $\theta_{min}$ values are clearly migrating toward the decreasing values time by 
time, and this migration repeats itself more than six times between orbital phases 0.0 and 1.0, through the 
4 years of long cadence data. In panel $a$, as the time progresses, $\theta_{min}$ values approach to 
$\theta_{min}=0.0$ and then jump to $\theta_{min}=1.0$. In order to see migration without interruption, 
one needs to add a proper integer to the $\theta_{min}$ values occurred before each jump, so that all 
$\theta_{min}$ values appear on a trend without 
any discontinuity. By this way, we obtain adjusted $\theta_{min}$ values (thus adjusted migration movement).
We plot the adjusted $\theta_{min}$ values versus Barycentric Julian Day in Figure~\ref{F9} panel $b$. 
Here, we note that a jump of 0.5 in phase is observed for the $\theta_{min}$ approximately around BJD 24\,55760. 
This behaviour is generally 
observed in the switching of the deeper minimum in the light curve by 0.5 in phase, which is commonly called 
as Flip-Flop in the literature \citep{Berdyugina_Jarvinen_2005AN}.

Using the least-squares method, we derived the best linear fit (Equation~\ref{Eq3}) for adjusted phases of 
the times of minima, where $\theta_{min}$ is an adjusted phase for the minimum time, while $\tau$ is the 
Barycentric Julian Date, taking $\tau_{0}$ as 24\,54954.633527 that corresponds to the time of the first 
minimum found from the residuals. The numbers in brackets in Equation~\ref{Eq3} denote the statistical error 
for the coefficients and the constant, for their last digits.

\begin{equation}
    \theta_{min}=-0.003772(6) \times [\tau-\tau_{0}]+6.782(5)
	\label{Eq3}
\end{equation}

With careful inspection of the Figure~\ref{F9} panel $b$, slight deviation of the adjusted $\theta_{min}$ 
values around the linear fit can be noticed. We plot the residuals from the linear fit in Figure~\ref{F9} 
panel $c$, which reveals a clear variation following two sine waves, one with a large amplitude 
and small frequency shown by a line in the figure, and another one with has small amplitude and high 
frequency.

\subsection{Flare activity and the OPEA model}\label{S3.6}

Apart from the rotational modulation of light curves caused by spot activity, both the long cadence and 
the short cadence data of KIC\,12418816 demonstrate that the system exhibits flare activity with high 
frequency. In this respect, first of all, using the residual data parts without any instant brightness increase 
were modelled by the Fourier Transform for each cycle. Then, the fit models derived from these Fourier Transforms 
were taken as the quiescent levels. Following the process described by \citet{Dal_Evren_2010AJ, Dal_Evren_2011a_AJ}, 
the start and end points were determined for each flare. Finally, all flare parameters were computed with respect to the
quiescent levels.

\begin{figure}
\centering
{\includegraphics[angle=0,scale=0.46,clip=true]{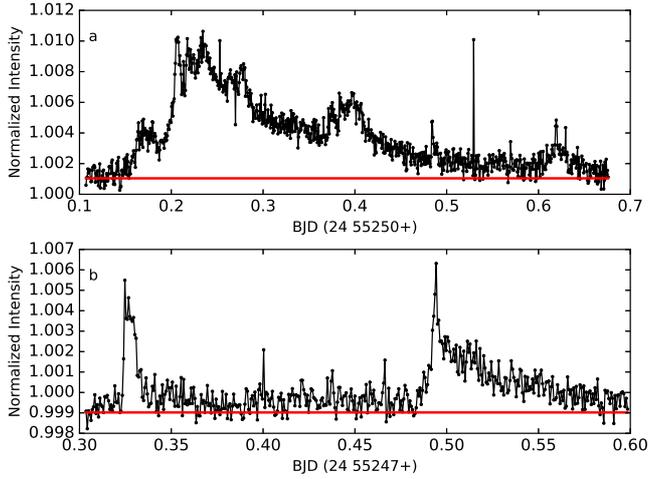}}
\caption{The flare light curve samples chosen from different parts of the short cadence data. 
In the figures, the filled circles represent the observations, while the red lines
represent the light level assumed as the quiescent state of the star.}
\label{F10}
\end{figure}

We detect 81 flares in total, where 73 of them are from the short cadence data, and 8 of them are from 
the long cadence data. We show two sample flares in Figure~\ref{F10}, which are detected from the short 
cadence data. For each flare, we first define beginning and end of the flare, and then compute the 
flare rise time ($\tau_{r}$), the decay time ($\tau_{d}$), amplitude of the flare maxima, and the 
flare equivalent duration ($P$). $P$ is defined as,

\begin{equation}
    P=\int(\dfrac{I_{flare}-I_{0}}{I_{0}})dt
	\label{Eq4}
\end{equation}

where $P$, $I_{0}$ and $I_{flare}$ are the flare-equivalent duration, the intensity of the star 
in the quiescent state and the intensity observed at the moment of the flare, respectively
\citep{Gershberg_1972Ap&SS}. We note that we do not compute the flare energies to be used in the 
following analyses, due to the reasons described in detail by \citet{Dal_Evren_2010AJ, Dal_Evren_2011a_AJ}. 
We tabulate computed parameters of 81 flares in Table~\ref{T5}.

\begin{table}
\setlength{\tabcolsep}{3pt}
\tiny
\caption{All the calculated parameters of flares detected from the short cadence observational data 
of KIC\,12418816. The flares are separated into two groups, as Group 1 and Group 2.}\label{T5}
\begin{center}
\begin{tabular}{ccccc}
\hline\noalign{\smallskip}
\multicolumn{5}{c}{Group1} \\
  Flare Time     &  $P$   & $\tau_{r}$ & $\tau_{r}$ &   Amplitude   \\
BJD (24 00000+)  &  (s)   &     (s)    &     (s)    &  (Intensity)  \\
\hline\noalign{\smallskip}
55209.95848	&	182.4619	&	1765.4980	&	10592.5540	&	0.063	\\
55246.31868	&	3.6229	&	411.9550	&	1176.9410	&	0.006	\\
55247.32470	&	4.9478	&	353.1170	&	1588.8960	&	0.007	\\
55248.05894	&	0.3060	&	58.9250	&	58.8380	&	0.000	\\
55249.52404	&	0.3441	&	117.7630	&	176.5150	&	0.003	\\
55249.88912	&	31.4420	&	176.5150	&	3060.1150	&	0.055	\\
55250.23445	&	123.1273	&	6944.2270	&	30366.0580	&	0.010	\\
55257.07293	&	1.7541	&	58.8380	&	117.6770	&	0.029	\\
55257.35491	&	4.5288	&	294.1920	&	1353.5420	&	0.006	\\
55257.67981	&	0.9500	&	58.8380	&	529.6320	&	0.002	\\
55258.50397	&	4.5736	&	176.6020	&	1588.8960	&	0.009	\\
55263.59266	&	3.2521	&	176.5150	&	1294.7040	&	0.006	\\
55269.47012	&	0.2125	&	58.8380	&	58.9250	&	0.003	\\
55269.49396	&	0.4468	&	117.6770	&	235.3540	&	0.005	\\
55271.04285	&	1.1811	&	176.5150	&	411.9550	&	0.005	\\
55273.88248	&	0.2896	&	176.5150	&	117.7630	&	0.003	\\
55274.62083	&	105.0486	&	529.8050	&	23010.1340	&	0.008	\\
55275.00771	&	1.5188	&	117.6770	&	706.2340	&	0.003	\\
55279.04444	&	249.9872	&	5296.4930	&	14123.9810	&	0.071	\\
55348.58201	&	103.8675	&	3531.0820	&	14124.1540	&	0.028	\\
55377.57815	&	125.6882	&	1765.4980	&	15889.5650	&	0.023	\\
55664.72893	&	171.9245	&	1765.4980	&	8827.4880	&	0.059	\\
55705.41346	&	223.9744	&	3531.0820	&	28248.3070	&	0.022	\\
55783.86009	&	258.5934	&	5296.4930	&	14123.8940	&	0.060	\\
55837.19165	&	444.8300	&	7061.7310	&	14123.4620	&	0.124	\\
\\
\multicolumn{5}{c}{Group2} \\
  Flare Time     &  $P$   & $\tau_{r}$ & $\tau_{r}$ &   Amplitude   \\
BJD (24 00000+)  &  (s)   &     (s)    &     (s)    &  (Intensity)  \\
\hline\noalign{\smallskip}
55247.49429	&	10.6828	&	706.1470	&	5355.2450	&	0.007	\\
55247.92544	&	2.2056	&	470.7940	&	1118.1020	&	0.005	\\
55248.01808	&	1.1038	&	353.0300	&	706.2340	&	0.002	\\
55248.17814	&	5.3484	&	1824.3360	&	5002.1280	&	0.001	\\
55248.27418	&	1.7568	&	353.1170	&	882.7490	&	0.004	\\
55248.36749	&	6.5505	&	1824.2500	&	3060.2020	&	0.002	\\
55248.52824	&	2.9749	&	294.2780	&	1647.7340	&	0.004	\\
55248.87425	&	2.0699	&	294.2780	&	1824.2500	&	0.000	\\
55249.75562	&	5.5426	&	353.1170	&	3707.5100	&	0.001	\\
55250.61928	&	7.6038	&	2883.6000	&	4884.4510	&	0.003	\\
55251.01569	&	3.9503	&	1353.5420	&	2353.9680	&	0.001	\\
55251.05179	&	4.4005	&	764.9860	&	1706.6590	&	0.005	\\
55251.22344	&	20.1882	&	5590.5980	&	10592.8130	&	0.003	\\
55251.35012	&	5.1587	&	353.1170	&	5473.0080	&	0.003	\\
55252.20017	&	15.8413	&	1824.3360	&	11181.2830	&	0.006	\\
55253.16395	&	6.1278	&	470.7940	&	3472.0700	&	0.003	\\
55254.07802	&	0.2145	&	58.8380	&	294.1920	&	0.000	\\
55254.10050	&	0.1640	&	58.8380	&	235.3540	&	0.000	\\
55254.28032	&	4.7232	&	470.7940	&	3472.0700	&	0.002	\\
55254.36069	&	1.0356	&	411.9550	&	588.4700	&	0.002	\\
55257.11652	&	1.3689	&	411.9550	&	647.3090	&	0.003	\\
55257.15194	&	1.0095	&	529.6320	&	353.1170	&	0.003	\\
55257.86848	&	4.4334	&	529.6320	&	3295.5550	&	0.004	\\
55257.91820	&	0.9397	&	294.2780	&	764.9860	&	0.002	\\
55258.48217	&	1.1849	&	353.1170	&	588.4700	&	0.004	\\
55258.66267	&	4.6701	&	1177.0270	&	2059.6900	&	0.003	\\
55258.80094	&	14.3333	&	1471.2190	&	7473.8590	&	0.003	\\
55258.92490	&	1.8988	&	588.4700	&	1412.3810	&	0.002	\\
55260.24220	&	1.9206	&	235.3540	&	1647.8210	&	0.002	\\
55260.45744	&	3.3707	&	294.2780	&	2412.8060	&	0.003	\\
55261.83195	&	2.0364	&	235.3540	&	1530.0580	&	0.002	\\
55261.92799	&	3.4647	&	1530.0580	&	2707.0850	&	0.002	\\
55263.44009	&	1.6343	&	411.9550	&	1118.1020	&	0.002	\\
55263.60901	&	0.7591	&	117.6770	&	470.8800	&	0.003	\\
55264.36642	&	2.1965	&	353.1170	&	1059.2640	&	0.004	\\
55264.59460	&	2.2930	&	353.0300	&	1059.3500	&	0.002	\\
55265.29071	&	2.5882	&	294.1920	&	1294.7040	&	0.004	\\
55265.31387	&	0.6201	&	176.5150	&	470.7940	&	0.002	\\
55266.82598	&	6.7981	&	1118.1890	&	2471.6450	&	0.005	\\
55267.17744	&	0.3133	&	176.5150	&	176.5150	&	0.003	\\
55267.18425	&	0.9339	&	411.9550	&	529.6320	&	0.003	\\
55267.35113	&	3.2132	&	235.4400	&	1471.2190	&	0.005	\\
55267.69169	&	11.2563	&	353.1170	&	6473.4340	&	0.005	\\
55268.44706	&	1.6236	&	411.9550	&	706.2340	&	0.003	\\
55268.54106	&	10.1567	&	4531.4210	&	3177.8780	&	0.003	\\
55270.06679	&	2.6880	&	176.5150	&	1235.8660	&	0.006	\\
55270.26091	&	3.2616	&	117.6770	&	2059.6900	&	0.003	\\
55270.29906	&	0.3316	&	58.8380	&	294.2780	&	0.002	\\
55270.76018	&	1.1611	&	58.8380	&	941.5870	&	0.004	\\
55270.80309	&	2.8562	&	470.7940	&	1942.0990	&	0.003	\\
55272.16331	&	5.2687	&	823.9100	&	2883.6860	&	0.003	\\
55272.20894	&	3.8012	&	1059.2640	&	2471.6450	&	0.002	\\
55273.01268	&	2.6030	&	294.1920	&	1294.7040	&	0.004	\\
55273.49832	&	7.3640	&	823.8240	&	11946.5280	&	0.006	\\
55273.77350	&	26.7146	&	6885.3890	&	8709.8110	&	0.003	\\
55274.54590	&	7.4994	&	2883.6000	&	4060.6270	&	0.002	\\
\hline
\end{tabular}
\end{center}
\end{table}

Examining the relationships between the flare parameters, it was seen that the distribution of 
flare equivalent durations on the logarithmic scale versus flare total durations are varying in 
a rule. The distribution of flare equivalent durations on the logarithmic scale cannot be higher 
than a specific value for the star, and it is no matter how long the flare total duration is. 
Using the SPSS V17.0 \citep{SPSS_Green_et_al_1996} and GrahpPad Prism V5.02 
\citep{Graphpad_motulsky2007} programs, \citet{Dal_Evren_2010AJ, Dal_Evren_2011a_AJ} demonstrated 
that the best function is the One Phase Exponential Association (hereafter $OPEA$) for the 
distributions of flare equivalent durations on the logarithmic scale versus flare total durations. 
The $OPEA$ function \citep{Graphpad_motulsky2007, Spanier_1987} has a Plateau term, and this makes 
it a special function in the analyses. The $OPEA$ function is defined by Equation~\ref{Eq5},

\begin{equation}
    y=y_{0}+(Plateau-y_{0}) \times (1-e^{-k \times x})
	\label{Eq5}
\end{equation}

where the parameter $y$ is the flare equivalent duration on a logarithmic scale, the parameter $x$ is 
the flare total duration as a variable parameter, according to the definition of \citet{Dal_Evren_2010AJ}. 
In addition, the parameter $y_{0}$ is the flare-equivalent duration on a logarithmic scale for the least 
total duration, which means that the parameter $y_{0}$ is the least equivalent duration occurring in a 
flare for a star. Logically, the parameter $y_{0}$ does not depend on only flare mechanism occurring on 
the star, but also depends on the sensitivity of the optical system used for the observations. In this case, 
the optical system is the optical systems of the $Kepler$ Satellite. The parameter Plateau value is the upper limit 
for the flare equivalent duration on a logarithmic scale. \citet{Dal_Evren_2011a_AJ} defined Plateau value 
as a saturation level for a star in the observing band.

After we derive the $OPEA$ model for all 81 flares, we notice that the correlation coefficient squared 
($R^{2}$) is very low, while the probability value ($p-value$) is very high. It means that the model does not 
perfectly fit the distributions. In fact, it seems that there are two arms in the distribution of flare equivalent 
durations on the logarithmic scale ($log~P$) versus flare total time ($\tau_{t}$) toward the higher equivalent durations. 
Particularly, this dissociation gets much clearer for the flares, whose total flare time is longer than 1700 s.
The flares with total flare times smaller than 1700 s seems that they mazily locate together in the 
distribution of flare equivalent durations on the logarithmic scale. In this point, we derived two independent models 
for the flares with total flare times larger than 1700 s. Then, following these independent model trends 
we split also the flares with total flare times smaller than 1700 s into two groups. Hence, 25 flares were grouped as Group 1, 
while the rest of them were grouped as Group 2.

The independent samples t-test (hereafter t-test) \citep{Dawson_Trap_2004basic, 2003psa..book.....W} was used in the SPSS V17.0 and GraphPad Prism V5.02 software in order to test whether these 
two groups are really independent from each other. The main average of the equivalent durations in the logarithmic scale 
for the flares in the plateau level of Group 1 was calculated and found to be 2.253$\pm$0.064 s with standard deviation of 0.203 s, and it was 
found to be 1.020$\pm$0.065 s with standard deviation of 0.222 s for the flares in the plateau level of Group 2. This analyses also confirmed that 
the separation between two groups is statistically acceptable.

\begin{figure}
\centering
\includegraphics[angle=0,scale=0.68,clip=true]{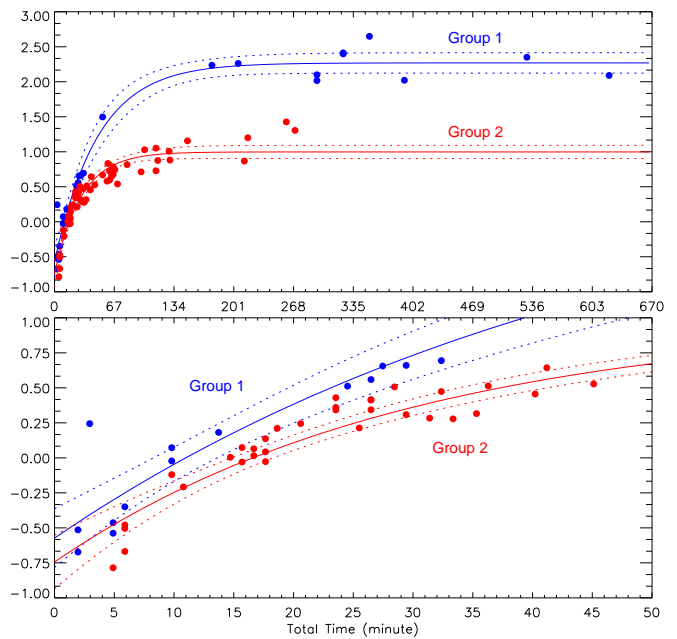}
\caption{Using the least-squares method, the OPEA models derived from the detected 81 flares. 
In the upper panel, the filled circles represent the observations, while the continuous lines and dashed lines 
show the models, and the confidence intervals of 95\%, respectively. The part of the figure, where the flares with 
low energy, are located is re-plotted to show the detail in this part.}
\label{F11}
\end{figure}

Finally, using the least-squares method, we derive the $OPEA$ model 
for each group, together with the confidence intervals of 95\%. Similarly, we split flares with the total times 
shorter than 1700 s into two groups, and derive $OPEA$ model of each group with the confidence intervals of 95\%. 
In the upper panel of Figure~\ref{F11}, we show distributions of flares in $\tau_{t}$ - $log~P$ 
plane, together with the confidence intervals of 95\%, while the flares with total flare times smaller than 1700 s 
are plotted in the bottom panel, zooming the axes. 
We list parameters of both models in Table~\ref{T6}, which are found by using the least-squares method. 
The span value listed in the table is difference between $Plateau$ and $y_{0}$ values. The half-life value is 
equal to $ln2/K$, where $K$ is a constant depending on a special $x$ value, where the model reaches to the 
Plateau value \citep{Dawson_Trap_2004basic}. In other words, the $n \times half-life$ parameter is the half 
of the minimum flare total time, which is enough to the maximum flare energy occurring in the flare mechanism.

\begin{table}
\tiny
\caption{The parameters obtained from the OPEA models, using the least-squares method.}\label{T6}
\begin{center}
\begin{tabular}{ccc}
\hline\noalign{\smallskip}
Parameter   &   Group1   &   Group2   \\
\hline\noalign{\smallskip}
$Y_{0}$       &   $-$0.5732$\pm$0.1013   &  $-$0.7440$\pm$0.0955  \\
$Plateau$     &     2.2692$\pm$0.0704    &    0.9975$\pm$0.0470   \\
$K$           &   0.00034$\pm$0.00004    &  0.00056$\pm$0.00006   \\
$\tau$        &           2926.2         &       1786.3           \\
Half-time     &           2028.3         &       1238.2           \\
Span          &    2.8425$\pm$0.1175     &   1.7415$\pm$0.0897    \\
\\
\multicolumn{3}{c}{95\% Confidence Intervals}                     \\
\hline\noalign{\smallskip}
$Y_{0}$       &  $-$0.78324 to -0.36321  &   $-$0.93572 to -0.55232 \\
$Plateau$     &     2.1233 to 2.4151     &    0.90322 to 1.0917    \\
$K$           &    0.00025 to 0.00043    &   0.00044 to 0.00068    \\
$\tau$        &    2316.9 to 3970.3      &    1462.2 to 2295.0     \\
Half-time     &    1605.9 to 2752.0      &    1013.5 to 1590.8     \\
Span          &    2.5987 to 3.0862      &    1.5614 to 1.9216     \\
\\
\multicolumn{3}{c}{Goodness of Fit}                     \\
\hline\noalign{\smallskip}
$R^{2}$       &         0.9682           &       0.9013      \\
$p-value$ (D'Agostino \& Pearson)     &   0.001      &        0.008   \\
$p-value$ (Shapiro-Wilk)              &   0.009      &        0.009   \\
$p-value$ (Kolmogorov-Smirnov)        &   0.001      &        0.001   \\
\noalign{\smallskip}\hline
\end{tabular}
\end{center}
\end{table}

We test the $OPEA$ models derived for both groups, by using three different methods, the 
D'Agostino-Pearson normality test, the Shapiro-Wilk normality test and the Kolmogorov-Smirnov 
test \citep{D'Agostino_1986book} to check whether there are any other functions to model 
the distributions on $\tau_{t}$ - $log~P$ plane. In these tests, the probability value 
($p-value$) was found to be smaller than 0.001, meaning that there is no other proper function to 
model the distribution \citep{Graphpad_motulsky2007, Spanier_1987}. Considering the correlation 
coefficient squared ($R^{2}$) values in the Talbe~\ref{T6}, we conclude that the separation of 
the flares into two groups is statistically real.

In the literature, the distribution of flare cumulative frequency ($\nu(P)$) called as 
"the flare energy spectrum", which is a frequency serial computed for different flare energy 
limits for a star, has been derived several decades ago to identify the character of the flare energy 
\citep{Gershberg_1972Ap&SS, Gershberg_2005}. The distribution of $\nu(P)$ has been derived by using 
Equation~\ref{Eq6} \citep{Gershberg_1972Ap&SS}. 

\begin{equation}
    \nu(P)=\int_{P_{min}}^{P_{max}} \nu(P)dP
	\label{Eq6}
\end{equation}

\begin{figure}
\centering
{\includegraphics[angle=0,scale=0.72,clip=true]{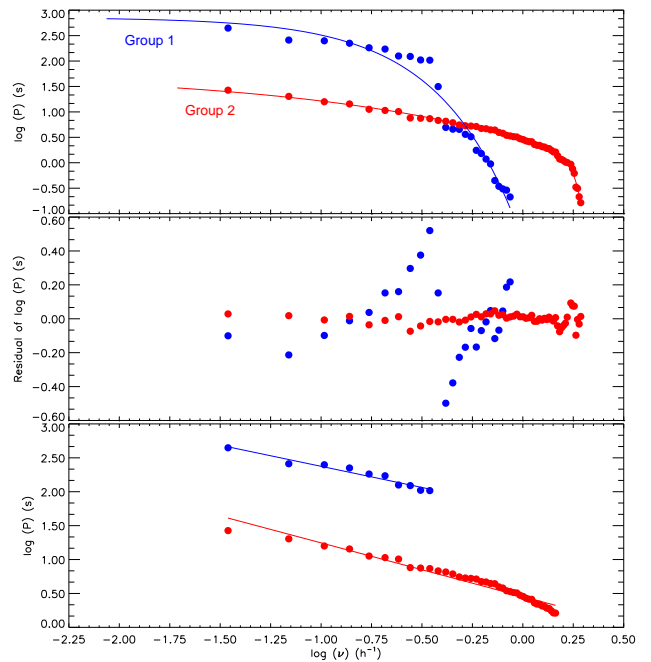}}
\caption{In the upper panel, the flare equivalent duration, $log~P$, distribution are plotted 
versus the flare cumulative frequency on the logarithmic scale for two groups of flares. The filled 
circles represent computed flare cumulative frequencies, while the lines show the exponential fits. 
In the middle panel, residuals from the exponential fits are shown. Colours have the same meaning 
as in the upper panel. In the bottom panel, linear parts of the distributions are plotted. 
In this panel, linear fits given by Equations~\ref{Eq7} and ~\ref{Eq8} are derived by using the 
least-squares method.}
\label{F12}
\end{figure}

However, the flare frequencies in the cumulative distribution have been computed for different limits 
of the flare equivalent durations instead of the flare energy, because of the luminosity term ($L$) in 
the energy equation of $E=P \times L$. The distribution of flare cumulative frequency is shown in 
Figure~\ref{F12} uppermost panel. We searched the best function to fit the distribution of flare cumulative 
frequency via least-squares method, and find that the most suitable function is an exponential function,
which provides correlation coefficient $R^{2}$ higher than 0.90. We test the possibility of any other
function, which might provide a better representation, with the methods applied to the OPEA model, and
find $p-values$ smaller than 0.001, indicating no alternative function to the exponential function. 
Figure~\ref{F12} middle panel shows residuals from the best-fitting exponential functions to the distributions.

\begin{equation}
    log(P)=-0.6234(\pm0.0493) \times 1.7510 (\pm 0.4238)
	\label{Eq7}
\end{equation}

\begin{equation}
    log(P)=-0.7931(\pm0.0260) \times 0.4526 (\pm 0.0117)
	\label{Eq8}
\end{equation}

The most important parameter is the slope of the linear fit derived for the linear part 
of the flare equivalent duration distribution plotted versus the flare cumulative frequency on the 
logarithmic scale \citep{Gershberg_2005}. In order to find the slopes for two groups, following 
\citet{Gershberg_2005}, the distribution were modelled by Equations~\ref{Eq7} and \ref{Eq8}. 
In a result, the slope of linear fit of Group 1 flares was found to be -0.6234$\pm$0.0493, while 
it was found to be -0.7931$\pm$0.0260 in the case of Group 2.

\begin{figure}
\centering
{\includegraphics[angle=0,scale=0.52,clip=true]{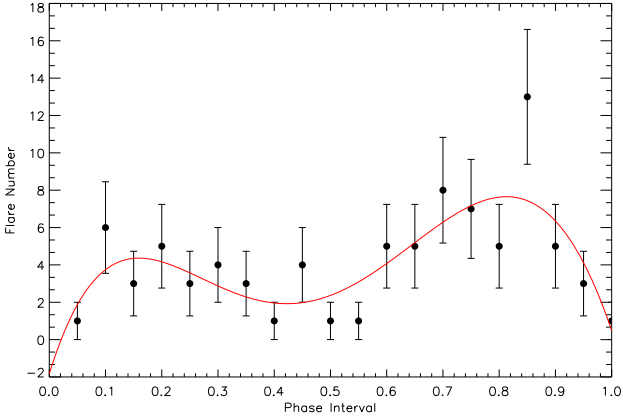}}
\caption{The distribution of flare total number in phase range of 0.05 is 
plotted with the simple Poisson error for each count versus phase for 81 flares.}
\label{F13}
\end{figure}

In order to find out where the flares occur on the active component, we derive the orbital phase distribution 
for all flares. In this respect, using the orbital period and ephemeris time given in Equation~\ref{Eq2}, we 
calculate the phase of each flare by considering the time at the moment of the flare. Figure~\ref{F13} shows 
the phase distribution of 81 flares, where the flare total number is computed for every 0.05 orbital phase step.

Among detected 81 flares, 25 of them belongs to the Group 1, while the remaining 56 flares belong to the Group 2.
\citet{Ishida_1991Ap&SS} described two frequencies for the stellar flare activity. These frequencies are defined 
as in Equation~\ref{Eq9} and Equation~\ref{Eq10},

\begin{equation}
    N_{1}=\Sigma n_{f}/\Sigma \tau_{t}
	\label{Eq9}
\end{equation}

\begin{equation}
    N_{2}=\Sigma P/\Sigma \tau_{t}
	\label{Eq10}
\end{equation}

where $\Sigma n_{f}$ is the total flare number detected in the observations, and $\Sigma \tau_{t}$ is the total 
observing duration, while $\Sigma P$ is the total equivalent duration obtained from all the flares. We compute both 
$N_{1}$ and $N_{2}$ flare frequencies for all flares. We carry out the same computation separately for both groups. 
Table~\ref{T7} tabulates the computation results.

\begin{table}
\tiny
\caption{Flare frequencies computed for all flares and grouped flares.}\label{T7}
\begin{center}
\begin{tabular}{cccrrrr}
\hline\noalign{\smallskip}
Parameters  &   All   &  Group 1    &    Group 2   \\
\hline\noalign{\smallskip}
Total Time (h)   &	693.91496    &   693.91496   &    693.91496    \\
Flare Number     &      81       &        25     &        56       \\
Total Equivalient Duration (s) &   2305.08277    & 2048.87285   &   256.20992  \\
$N_{1}$ ($h^{-1}$)  &    0.11673   &    0.03603   &    0.08070  \\
$N_{2}$ &    0.00092   &    0.00082   &    0.00010  \\
\noalign{\smallskip}\hline
\end{tabular}
\end{center}
\end{table}

\section{Discussion and results}

Spectroscopic and photometric analysis of KIC\,12418816 revealed that the system is a detached binary 
on a circular orbit. Combined spectroscopic orbit and light curve model parameters lead to the 
0.88$\pm$0.06 \Msun of mass, 0.85$\pm$0.02 \Rsun of radius and 4.521$\pm$0.015 of log $g$ for the
primary component, while the same parameters are 0.84$\pm$0.05 \Msun, 0.84$\pm$0.02 \Rsun and 
4.511$\pm$0.009 for the secondary component. Calculated composite spectrum of the system via 
54\,Psc + 54\,Psc configuration provides very good agreement with the observed composite spectrum 
of the system. Moreover, we observe strong emission from both components in the core of 
\ion{Ca}{ii} H\&K lines, while $H_{\beta}$ and $H_{\alpha}$ lines of both components are in form of 
filled emission, and almost embedded into the continuum level. These emission features clearly 
indicate strong chromosperic activity of both components. We also observe the traces of the activity 
of the components both in the $O-C$ diagram as irregular sinusoidal wave pattern, and in the light curves 
as frequent flares and light variation at out-of-eclipse. All These findings indicate that the system is 
composed of two almost twin and very active K0 main sequence stars with a detached configuration, 
which makes the system unique among its analogues. 

The age of the system is given as log(age)=9.53 Gyr by \citet{Morton_et_al_2016ApJ}. Plotting the 
components on $Log~T_{eff}-Log~L/L_{\odot}$ plane, we find that the components are located close to the
Zero Age Main Sequence (ZAMS) rather than the Terminal Age Main Sequence (TAMS), thus indicating that
both components must be relatively younger than the age reported in \citet{Morton_et_al_2016ApJ}.

The light variation at out-of-eclipse phases indicates an unstable sinusoidal variation with a variable
amplitude from one cycle to the next, while $\theta_{min}$ values migrates towards the decreasing phases.
Considering the properties of the components, rotational modulation of cool spots is the most possible 
explanation for the out-of-eclipse variations. However, it is hard to decide which component is actually 
spotted since both components exhibit emission features in \ion{Ca}{ii} H\&K, $H_{\beta}$ and $H_{\alpha}$
lines. Assuming that both components possess cool spots on their surface, one would expect a complicated 
light curve pattern at out-of-eclipse phases, due to the interfering rotational modulations from both 
components. Furthermore, $\theta_{min}$ values would have a scattered distribution versus time. In the 
case of KIC\,12418816, we observe rather regular patterns at out-of-eclipse phases, which resembles typical 
light curve of a spotted single star. We observe the similar regularity in the distribution of $\theta_{min}$ 
values (Figure~\ref{F9}), which smoothly follows a linear trend. These suggest that only one of the component 
is spotted. Inspecting the strength of emission (Figure~\ref{F4}), one can see that the emission strength of 
the secondary component in \ion{Ca}{ii} H\&K, $H_{\beta}$ and $H_{\alpha}$ lines seems generally stronger 
compared to the primary component, thus the secondary component is possibly the spotted component. In this case, 
the primary component still exhibits chromospheric activity, but without rotational modulation of spots. We note
that both components have flare activity, no matter whether they have spots on their surface. There are several 
stars exhibiting flare activity without any rotational modulation in their light curves, as in the case of the
primary component of KIC\,12418816. AD\,Leo and EQ\,Peg are such stars, which are well-known UV\,Ceti type 
variables \citep{Dal_Evren_2011a_AJ, Dal_Evren_2011b_PASJ}.

We investigated migration behaviour of the spotted areas by tracing $\theta_{min}$ values, and find that
$\theta_{min}$ values migrates towards decreasing phases as the time progress. Migration between 0.99
and 0.00 phases repeats itself $\sim$6.5 times through the 4 years of $Kepler$ photometry and obviously
indicate longitudinal migration of active region on the surface of the active (secondary) component. Applying
linear fit to the adjusted $\theta_{min}$ values, we find the migration period of the active region as 
0.72$\pm$0.05 year. This period is similar to the solar analogue stars. Using the orbital period and the migration 
period of KIC\,12418816, one can find mean rotation period of the spotted component, as described by 
\citet{Hall_Busby_1990_difrot}, and estimate the surface share via the difference between the mean rotation 
period and the orbital period. We find the mean rotation period as $1.51\pm0.21$ day, which leads to the surface
share $\Delta \Omega=0.02\pm0.87$ rad day$^{-1}$. Since the statistical uncertainty is very large compared to
the shear value, we refrain further discussion on the surface shear.

Another remarkable spot activity property of KIC\,12418816 comes from the $\theta_{min}$ shift versus time. 
The residual of $\theta_{min}$ shift obtained after the linear correction exhibits a systematic 
sinusoidal variation, in which there is also a quasi-periodic wave like second variation with smaller 
amplitude. The systematic sinusoidal variation with larger amplitude should be caused by globally swinging 
position of spotted area due to differential rotation on the stellar surface. However, the quasi-periodic 
wave like second variation with smaller amplitude should be affected by local spots which rapidly evolve on 
the stellar surface. Figure~\ref{F8} panel $b$ provides clear evidence of such a rapid spot evolution, and
migration.

Apart from the spot activity, the most notable variation is flare activity at out-of-eclipses. After 
determining flare parameters of 81 detected flares, we modelled the distribution of flare equivalent durations on the 
logarithmic scale versus the flare total time with the OPEA model (Equation~\ref{Eq5}). The initial attempts did not 
give any statistically acceptable model for the distribution due to large scattering. The similar phenomenon 
is seen in the flare study of \citet{Kamil_Dal_2017PASA}. Considering both $R^{2}$ and the $p-values$ derived for the 
model, they modelled the flare equivalent duration data with two different OPEA models. Following the way they defined, 
we modelled the flare equivalent duration data with two different OPEA models in this study. We found plateau value as
2.2692$\pm$0.0704 s from the model of Group 1 flares, while it was computed as 0.9975$\pm$0.0470 s from the model of 
Group 2. The plateau value of first group is 2 times larger than that found from Group 2.

The main average of the equivalent durations in the logarithmic scale 
for the plateau flares of Group 1 was found to be 2.253$\pm$0.064 s with standard deviation of 0.203 s, and it was 
found to be 1.020$\pm$0.065 s with standard deviation of 0.222 s for Group 2. 
The results confirm that these two groups are statistically different from each other.

In this point, one may claim that the separation is caused due to the flare morphology 
\citep{1974ApJS...29....1M, 2010AJ....140..483D, 2014ApJ...797..121H, 2014ApJ...797..122D}. For example, one can claim 
that slow flares are aggregated in Group 2, while the fast flares are aggregated in Group 1. In fact, according to 
\citet{2010AJ....140..483D, 2014ApJ...797..121H, 2014ApJ...797..122D}, a slow flare has lower energy than a fast flare 
with same duration. Therefore, this situation can explain the 2 times difference between these two groups. 
To test whether there is any morphological affect, the light variations of all flares were morphologically checked, 
and we see that both groups have both flare morphologies.

However, according to \citet{Dal_Evren_2010AJ, Dal_Evren_2011a_AJ}, the plateau value is defined 
as a saturation level of white-light flares for a star, and also a star can have just one plateau values depending 
on its $B-V$ color index. Thus, if the plateau values are markedly different for two group flares, it means that 
these flare groups come from two different stars, whose $B-V$ color indexes are different from each other.
\citet{Dal_2012PASJ} showed that the plateau values systematically vary depending on the $B-V$ colours. In this case, 
the flares compiled as Group 1 and Group 2 in this study must come from two different sources. The magnetic activity
sensitive lines of both two components exhibit strong emission. These two results confirm each other, i.e. according 
to spectral data, both components have high level magnetic activity, therefore both of them must exhibit flare activity.

In the literature, there is no relation between flare saturation level and existence of stellar cool spot activity. 
For instance, in the case of KOI-256, one of the interesting binary system observed by $Kepler$ Satellite, 
the plateau value was found to be 2.3121$\pm$0.0964 s \citep{yoldas_dal_2017}. In the case of FL\,Lyr, it was found 
to be 1.232$\pm$0.069 s \citep{Yoldas_Dal_2016PASA}. The flares compiled as Group 1 in this study behave as the same 
as flares detected from KOI-256, while Group 2 flares behave as the same as flares detected from FL\,Lyr. 
\citet{Yoldas_Dal_2016PASA, yoldas_dal_2017} revealed that both KOI-256 and FL\,Lyr have high level spot 
activity on their surfaces. Comparing these samples, one expects that both components of KIC\,12418816 could exhibit 
spot activity depending on their plateau values. However, our findings above indicate that just one of the component
possesses spotted areas on its surface. A similar case has been described by \citet{Kamil_Dal_2017PASA} for 
KIC\,2557430. In the case of KIC\,2557430, although there are two sources exhibiting flares with different plateau 
levels, only one of them exhibits stellar spot activity.

The mostly discussed model and parameter in the literature are the cumulative flare frequency distribution 
and the slope of its linear fit \citep{Gershberg_1972Ap&SS, Lac76, Wal81, Gershberg_Shakhovskaia_1983Ap&SS, Pet84, 
Mav86, Gershberg_2005, Haw14, Dav14}. In this study, the cumulative flare frequency distributions were derived for 
the flares in both groups. Modelling the distributions by Equations~\ref{Eq7} and \ref{Eq8}, the linear fits were 
derived. The slope of linear fit of Group 1 flares was computed as -0.6234$\pm$0.0493, while it was computed as 
$-$0.7931$\pm$0.0260 for Group 2 flares.

Besides the plateau value, there is one more indicator for the flare activity level of stars. \citep{Ishida_1991Ap&SS} 
described $N_{1}$ flare frequencies. $N_{1}$ flare frequency given by Equation~\ref{Eq9} is the flare number 
per hour, while $N_{2}$ flare frequency given by Equation~\ref{Eq10} is the flare total equivalent duration emitting 
per hour. We find $N_{1}$ as 0.03603 $h^{-1}$ for flares of Group 1 and 0.08070 $h^{-1}$ for Group 2. It means that the 
Group 2 flares are more frequent than Group 1 flares. However, as it is seen from $N_{2}$ flare frequencies listed 
in Table 7, the flares of Group 1 are more powerful than those of Group 2. In a result, Group 1 flares are rare but
powerful, while Group 2 flare are less powerful compared to the Group 1 flares but more frequent.

In general, an unexpected result comes from the orbital phase distribution of flares. Unlike what is known 
and discussed in the literature \citep{Haw14}, the flare number seems to be rising around the phases of 0.15 and 0.85
(Figure~\ref{F13}). It means that one can observe the flares more frequently just before and just after 
the primary eclipse of the whole light curve. This is very impressive because it indicates some clue about magnetic
interaction between the components. Owing to the orbital properties of the system, The components are close enough to 
each other to get an interaction easily when a flare occurs on surface of a component. The situation makes the system an
interesting binary to study for the readers working on magnetic natures of binaries in different evolutionary stages.

\section*{Acknowledgements}

We wish to thank the Turkish Scientific and Technical Research Council (T\"UB\.ITAK) for supporting this work 
through grant No. 116F213, and for a partial support in using RTT150 (Russian-Turkish 1.5-m telescope in Antalya) 
with project number 14BRTT150-667. We also thank the referee for useful comments that have contributed to the improvement of the paper.





\bsp	
\label{lastpage}
\end{document}